\documentclass[11pt,aps,nofootinbib,floatfix,prd,tightenlines]{revtex4}%
\usepackage{amsmath}
\usepackage{amsfonts}
\usepackage{amssymb}
\usepackage{graphicx}
\usepackage{graphics}
\usepackage{epsfig}
\usepackage[section]{placeins}
\usepackage{multirow}
\usepackage{bibentry}
\usepackage{braket}
\usepackage[font=footnotesize,labelfont=bf]{caption}
\usepackage{pdfpages}
\usepackage{slashed}
\usepackage{bbm} %matrix identity with \mathbbm{1}
\providecommand{\U}[1]{\protect\rule{.1in}{.1in}}
\nobibliography*

\def\){\right)} 
\def\({\left(} 
\def\]{\right]} 
\def\[{\left[}

\begin{document}

\title{Radiative corrections to elastic muon-proton scattering\\ at low momentum transfers}

\author{Norbert Kaiser}
\email{nkaiser@ph.tum.de}
\affiliation{Physik-Department T39, Technische Universit\"{a}t M\"{u}nchen,
   D-85748 Garching, Germany}

\author{Yong-Hui Lin}
\email{yonghui@hiskp.uni-bonn.de}
\affiliation{Helmholtz-Institut f{\"u}r Strahlen- und Kernphysik and Bethe Center
  for Theoretical Physics, Universit\"at Bonn, D-53115 Bonn, Germany}

\author{Ulf-G. Mei{\ss}ner}
\email{meissner@hiskp.uni-bonn.de}
\affiliation{Helmholtz-Institut f{\"u}r Strahlen- und Kernphysik and Bethe Center
for Theoretical Physics, Universit\"at Bonn, D-53115 Bonn, Germany\\%}
%\affiliation{
Institute~for~Advanced~Simulation, Institut~f\"{u}r~Kernphysik,
Center for Advanced Simulation and Analytics, and
J\"{u}lich~Center~for~Hadron~Physics, \\ 
Forschungszentrum~J\"{u}lich, D-52425~J\"{u}lich,~Germany\\%}
%\affiliation{
Tbilisi State University, 0186 Tbilisi, Georgia}

\begin{abstract}
We systematically calculate the radiative corrections of order $\alpha/\pi $ to
elastic muon-proton scattering at low momentum transfers. These include vacuum
polarization, photon-loop form factors of the muon and the proton, two-photon
exchange corrections and soft photon radiation. In particular, we discuss these
corrections for the kinematics of the upcomimg AMBER experiment with a $100\,$GeV
muon beam energy. It is found that for the ratio to the Born cross section, only
the minor terms from the photon-loop form factors of the proton and two-photon
exchange depend on the proton structure predetermined by the strong interactions.
Since a prominent role among the radiative corrections is played by soft photon
radiation, the calculation of the bremsstrahlung process $\mu p \to\mu p \gamma$
should be extended beyond the soft photon approximation and tailored to the
specific experimental conditions.
\end{abstract}

%\date{\today}

\maketitle

\vfill

\pagebreak

\section{Introduction}
\label{sec:intro}

Elastic muon-proton scattering at low momentum transfers offers an alternative
method to measure the proton charge radius $r_p$, which is a fundamental
quantity in the theory of the strong interactions. It is defined here by  the
slope of the proton charge form factor $G_{E,p}(Q)$ at zero momentum transfer
$r_p^2 = -6\, {dG_{E,p}(Q)}/{dQ^2}|_{Q=0}$, with $Q^2$ the invariant four-momentum transfer squared.
Any deviation from the value measured in electron-proton scattering would challenge the concept
of lepton-flavor universality, which is a cornerstone of the so successful Standard Model of particle
physics, that has been challenged in recent experiments on certain decay modes of B mesons,
see Ref.~\cite{Bifani:2018zmi} for a recent review.
Two experiments are pursuing such proton radius measurements, namely MUSE at PSI \cite{Downie:2014qna} and
AMBER at CERN \cite{Adams:2018pwt}. Both experiments were triggered by the so-called ``proton radius puzzle'',
see e.g. Ref.~\cite{Pohl:2013yb}, but it must be said that most recent determinations of the proton
radius from electron-proton scattering and the Lamb shift in electronic hydrogen are in favor of the so-called
small radius, $r_p \simeq 0.84\,$fm, as collected in Tab.~\ref{tab:rp}. The small value is further supported by
a dispersion-theoretical analysis of all existing scattering and annihilation data in the space-like and the
time-like region~\cite{Lin:2021xrc},
as also shown in the table. The underlying dispersive framework and the history of proton radius extractions
based on dispersion relations (DRs) are discussed in detail in Ref.~\cite{Lin:2021umz}. Still, an independent
extraction from $\mu ^\mp p$ scattering would be highly welcome, further complementing the groundbreaking
work on the Lamb shift in muonic hydrogen~\cite{Pohl:2010zza}, which essentially initiated the whole
proton radius discussion.

\begin{table}[h!]
\caption{Modern precision extractions of the proton charge radius $r_p$ from
  the Lamb shift in electronic hydrogen and electron-proton scattering as
  well as dispersion theory.\label{tab:rp}}
\begin{center}
\begin{tabular}{|c|c|c|c|}
\hline
$r_p$ [fm] & year &  method & Ref. \\
\hline
%0.8335(95) & 2017 & el.  Lamb shift &\cite{Beyer:2017gug}\\
0.877(13)  & 2018 & H Lamb shift &\cite{Fleurbaey:2018fih}\\
0.833(10)  & 2019 & H Lamb shift &\cite{Bezginov:2019mdi}\\
0.8482(38) & 2020 & H Lamb shift &\cite{Griffin:2020}\\
0.8584(51) & 2021 & H Lamb shift &\cite{Brandt:2021yor}\\
0.831(7)(12) & 2019& $ep$ scattering & \cite{Jlab}\\
\hline
0.840(3)(2) & 2022& disp. theory & \cite{Lin:2021xrc}\\
\hline
\end{tabular}
\end{center}
\vspace{-3mm}
\end{table}  

In this work we consider the radiative corrections to $\mu^\mp p$ scattering specifically
for the kinematics of the AMBER experiment, which operates with a high-energetic muon beam at $100\,$GeV
and measures in near forward directions, thus spanning the momentum transfers  $32\,$MeV$\,< Q < 141\,$MeV,
which nicely overlaps with the range of the upcoming MUSE experimemnt at PSI with $45\,$MeV$\, < Q< 265$~MeV, 
the PRAD-II experiment at Jefferson Lab for $e^-p$ scattering with $14\,$MeV$\, < Q< 245$~MeV~\cite{PRad:2020oor}
as well as the MAGIC $e^-p$ experiment at Mainz, that aims at a momentum range
$10\,$MeV$\,<Q<292$~MeV~\cite{Denig:2016tpq}.
The AMBER experiment intends to measure the proton radius with an accucary of better than $0.01$~fm,
which requires a detailed study of the radiative corrections to be able to achieve such an accuracy.
Such a calculation is provided here, based on the works in Refs.~\cite{Kaiser:2010zz,Kaiser:2016tbf}
employing the best phenomenological available proton form factors from Ref.~\cite{Lin:2021xrc}.
For related work on radiative corrections to muon-proton scattering,
see Refs.~\cite{Tomalak:2015hva,Tomalak:2018jak,Peset:2021iul}.

The manuscript is organized as follows: In Sec.~\ref{sec:diffXS} we display the differential
cross section for $\mu^- p$ scattering including the various radiative correction terms. These are
discussed in detail in the following sections, namely the  photon-loop form factors of the muon and of
the proton in Sec.~\ref{sec:phloop}, the two-photon corrections in Sec.~\ref{sec:tpe} and the
soft-photon radiation in Sec.~\ref{sec:softrad}. Finally, in Sec.~\ref{sec:res}, we put all pieces
together and display and discuss the radiative corrections for the AMBER kinematics. We end
with a short summary and an outlook in Sec.~\ref{sec:summ}.

\section{Differential cross section}
\label{sec:diffXS}

We consider elastic muon scattering off protons, specifically the process $\mu^-(k_1)+ p(p_1)\to \mu^-(k_2)+p(p_2)$,
and introduce the dimensionless Mandelstam variables 
\begin{equation}
  s=(p_1+k_1)^2/M^2\,, \quad t = (k_1-k_2)^2/M^2\,, \quad  u= (p_1-k_2)/M^2\,,
\end{equation}  
that satisfy the constraint $s+t+u = 2+2r$, with $M=938.272\,$MeV the proton mass and the squared muon-to-proton
mass ratio $r= (m_\mu/M)^2 = 1.2681\cdot 10^{-2}$. The advantage of these (uncommon) dimensionless variables
$(s,t,u)$ is that they allow us to write the differential cross section and radiative corrections in concise analytical
form without repeating  permanently the mass parameters.

The unpolarized differential cross section for $\mu^- p\to \mu^- p$ including radiative corrections of order
$\alpha/\pi$, with $\alpha = 1/137.036$ the electromagnetic fine-structure constant, reads:
\begin{equation}
  {d\sigma \over dt} = {4\pi \alpha^2 \over M^2t^2P } \Big\{ H_0\big(1+ 2 \Pi_{\rm vp}+\delta_{\rm soft}\big) +H_1
  + H_2\Big\}\,,
  \label{eq:dxs}
\end{equation}
where the polynomial $P=s^2-2s(1+r)+(1-r)^2$  is equal  to the K\"all\'en function $\lambda(s,1,r)$. The first
term proportional to $H_0$ gives the Rosenbluth formula, generalized by the inclusion of the finite lepton mass,
and reads:
\begin{eqnarray}
  H_0 &\!\!\!=\!\!\!& \bigg[{(s+1-r)^2 \over 4-t}+r-s\bigg]\big( 4 G_E^2-t G_M^2\big)
  +t\Big(r+{t\over 2}\Big) G_M^2 \,.
\end{eqnarray} 
The argument $Q=\sqrt{-t} M$ of the proton electric and magnetic form factors $G_{E}(Q)$  and $G_{M}(Q)$, respectively,
that arise from the non-perturbative strong interactions, is not displayed explicitely.
The second term in Eq.~\eqref{eq:dxs} describes vacuum polarization in the one-photon exchange through the $Q$-dependent
function $\Pi_{\rm vp}$. For the low momentum transfers considered in this work, vacuum polarization due to the
two lightest leptons is sufficient:
\begin{equation}
\Pi_{\rm vp} = {\alpha \over 3\pi} \sum_{e,\mu} \bigg[ {1\over \tau^2} -{5\over 3} +{2\tau^2 -1 \over \tau^3}
\sqrt{1+\tau^2}\ln \big( \tau+\sqrt{1+\tau^2}\,\big)\bigg]\,,
\end{equation}
where $\tau_e =Q/2m_e$ and   $\tau_\mu =Q/2m_\mu$.  At $Q=300\,$MeV one gets $\Pi_{\rm vp} =(0.86+0.07)\% = 0.93\,\%$,
and at  $Q=100\,$MeV one has $\Pi_{\rm vp} =(0.69+0.01)\% = 0.70\,\%$. These numbers demonstrate the dominance
of electronic vacuum polarization. Further features are illustrated in Fig.~82 of
Ref.~\cite{WorkingGrouponRadiativeCorrections:2010bjp}, which shows the quantity $|1+\Pi_{\rm vp}|^2$ in
the space-like and time-like regions below 2\,GeV and thereby delineates the region, where leptonic
vacuum polarization is actually dominant.

The next correction factor $\delta_{\rm soft}$ arises from soft photon bremsstrahlung and its derivation will
be outlined in Sec.~\ref{sec:softrad}, starting from the basic soft photon amplitude. Moreover, the term proportional to 
$H_1$ in Eq.~\eqref{eq:dxs} gives twice the interference term of one-photon exchange with electromagnetic vertex
corrections  at the muon and at the proton. It is given by the expression
\begin{eqnarray}
\label{eq:H1}
H_1 &\!\!\!=\!\!\!& 2F_1^\gamma  H_0 +F_2^\gamma\, t\big(2G_E^2+t G_M^2\big)\nonumber \\
&& + 8 G_E^\gamma \bigg[{(s+1-r)^2 \over 4-t}+r-s\bigg]G_E+ G_M^\gamma \bigg[2s+t- {2(s+1-r)^2 \over 4-t}\bigg]t G_M\,, 
\end{eqnarray}
where $F_{1,2}^\gamma$ denote the photon-loop form factors of the muon and $G_{E,M}^\gamma$ are those of the proton.
These quantities are discussed in some detail in Sec.~\ref{sec:phloop}.

Finally, the last piece proportional to $H_2$ in Eq.~\eqref{eq:dxs} gives twice the interference term of one-photon exchange with the planar
and crossed two-photon exchange box diagrams, see Sec.~\ref{sec:tpe}. Note that $H_0$ and $H_1$ are even under 
$s \leftrightarrow u$, while $H_2$ is odd. The differential cross section for  the process $\mu^+(k_1)+ p(p_1)\to \mu^+(k_2)+p(p_2)$ is thus obtained by  $s \leftrightarrow u$ crossing of the terms in the curly brackets of Eq.~\eqref{eq:dxs}, i.e.,
one merely has to change the sign of $H_2$.

\section{Photon-loop form factors of muon and proton}
\label{sec:phloop}

\begin{figure}[t!]
\centering \includegraphics[width=8.cm,clip]{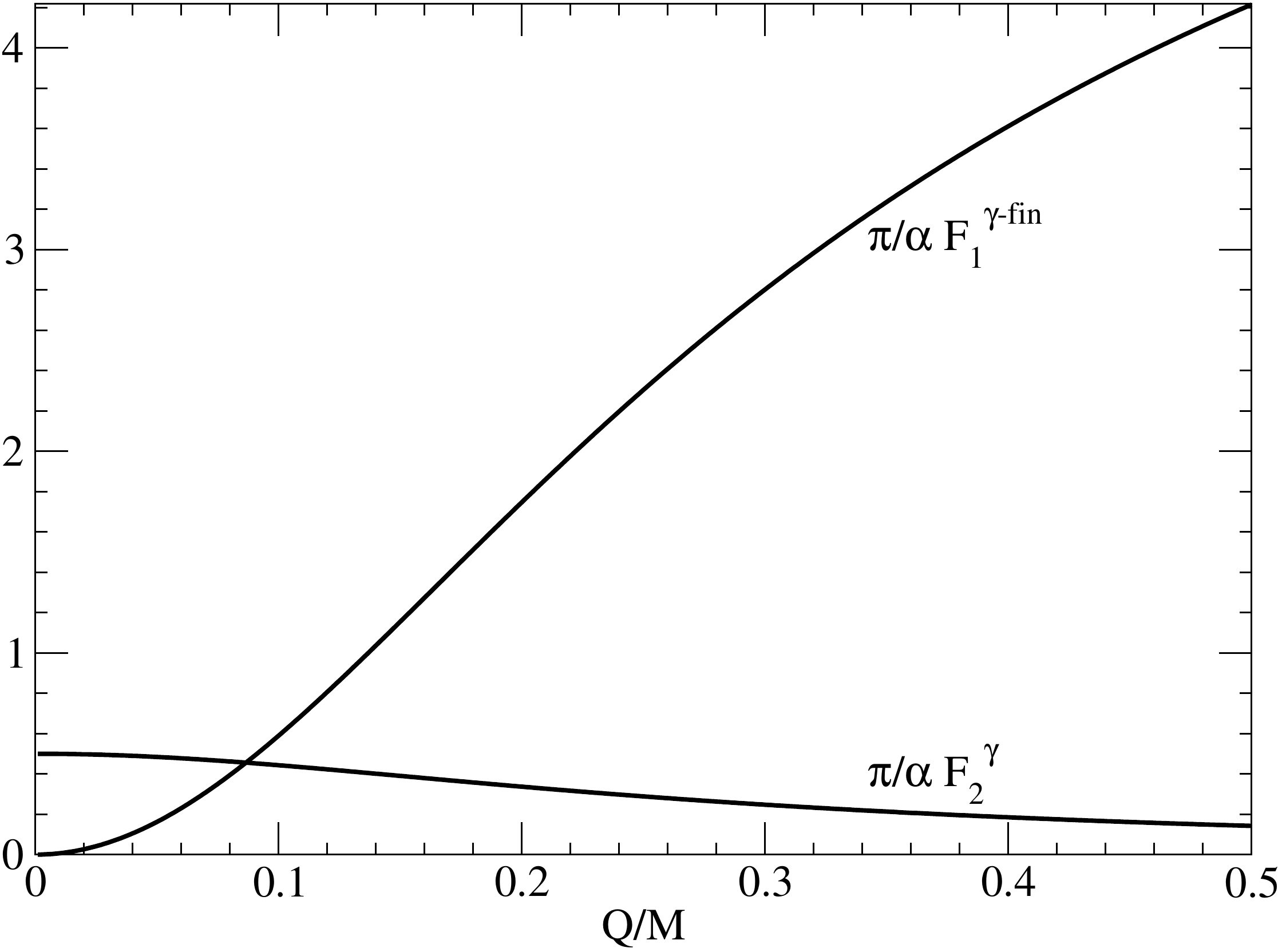}
\caption{Photon-loop form factors $F_{1,2}^\gamma$ of the muon,
where $F_1^{\gamma-{\rm fin}}$ refers to the infrared-finite part.}
\label{fig:muff}
\end{figure}

In this section, we discuss the photon-loop form factors of the muon and of the proton that appear in the expression for 
$H_1$, see Eq.~\eqref{eq:H1}.
The photon-loop form factors of the muon are obtained from a standard QED calculation of the triangle diagram
(and self-energy diagram which contributes through the muon wavefunction renormalization factor $Z_2$) and they
explicitly read:
%\begin{eqnarray}F_1^\gamma &\!\!\!=\!\!\!& {\alpha \over 2\pi }\bigg\{ (2\xi_{\rm ir}+\ln r)\bigg( 1+{2t-4r\over \sqrt{t^2-4r t}} \ln{\sqrt{4r-t}+\sqrt{-t} \over 2\sqrt{r}}\bigg)-2 \nonumber \\ &&  +{8r-3t\over  \sqrt{t^2-4r t}} \ln{\sqrt{4r-t}+\sqrt{-t} \over 2\sqrt{r}} +(t-2r) \int_4^\infty\! dx {\ln(x-4) \over(x r -t)\sqrt{x^2-4x}}\bigg\} \,, \end{eqnarray}
\begin{eqnarray}
\label{eq:F1}  
F_1^\gamma &\!\!\!=\!\!\!& {\alpha t \over 2\pi}  \int_4^\infty\! dx{ 1\over x(x r -t)\sqrt{x^2-4x}}
\bigg\{\big[2\xi_{\rm ir}+\ln r+ \ln(x-4)\big](x-2)+4 -{3x \over 2}\bigg\}  \nonumber \\
&\!\!\!=\!\!\!& {\alpha \over \pi }\bigg\{ (2\xi_{\rm ir}+\ln r)\bigg( {1\over 2}+{t-2r\over \sqrt{t^2-4r t}}
\ln{\sqrt{4r-t}+\sqrt{-t} \over 2\sqrt{r}}\bigg)+ {(2r-t)\over \sqrt{t^2-4rt}} \bigg[ 
\ln^2{\sqrt{4r-t}+\sqrt{-t} \over 2\sqrt{r}}  \nonumber \\ && +\bigg( {8r-3t\over 4r-2t}  -\ln{4r-t\over r}\bigg)
\ln{\sqrt{4r-t}+\sqrt{-t} \over 2\sqrt{r}} +{\rm Li}_2\bigg( {t-2r+\sqrt{t^2-4rt}\over 2r}\bigg)
+{\pi^2 \over 12}\bigg]-1 \bigg\} \,,\\
F_2^\gamma &=& {2 \alpha \over \pi }{r\over \sqrt{t^2-4r t}} \ln{\sqrt{4r-t}+\sqrt{-t}
  \over 2\sqrt{r}}\,.
\label{eq:F2}
\end{eqnarray}
The infrared divergence $\xi_{\rm ir}$ is defined as $\xi_{\rm ir} = \ln(M/m_\gamma)$, with $m_\gamma$ an
infinitesimal regulator photon mass. Moreover, Li$_2(a) = a \int_1^\infty \!dx [x(x-a)]^{-1}\ln x$ denotes
the conventional dilogarithmic function for $a<1$. One observes that the $F_{1,2}^\gamma$ are actually functions
of the ratio $-t/r$ and the first version of $F_1^\gamma$ is a once-subtracted dispersion relation.  Note that
$F_1^\gamma(0) =0$ and  $F_2^\gamma(0) = \alpha/2\pi = 1.1614\!\cdot\!10^{-3}$ gives the leading
correction to the muon anomalous magnetic moment. These form factors are shown (multiplied with $\pi/\alpha$) in Fig.~\ref{fig:muff}, leaving out the regularization-dependent $\xi_{\rm ir}$ term for $F_1^\gamma$.

The photon-loop form factors of the proton are composed in a similar way of infrared-divergent and
infrared-finite pieces:
\begin{equation}
G_{E,M}^\gamma = {\alpha \over \pi }\xi_{\rm ir}\bigg( 1+{2t-4\over \sqrt{t^2-4 t}} \ln{\sqrt{4-t}+\sqrt{-t} \over 2}
\bigg)G_{E,M}^{} +  G_{E,M}^{\gamma{\rm-fin}}+G_{E,M}^{\gamma\Delta}+G_{E,M}^{\gamma\Delta\Delta}\,, 
\end{equation}
where $G_{E,M}^{\gamma{\rm-fin}}$ denotes the infrared-finite part and $G_{E,M}^{\gamma\Delta}+G_{E,M}^{\gamma\Delta\Delta}$
arises from treating inelastic contributions through single and double $\Delta^+(1232)$ resonance excitation
of the proton.  

\begin{figure}[t!]
\centering
\includegraphics[width=0.4\textwidth]{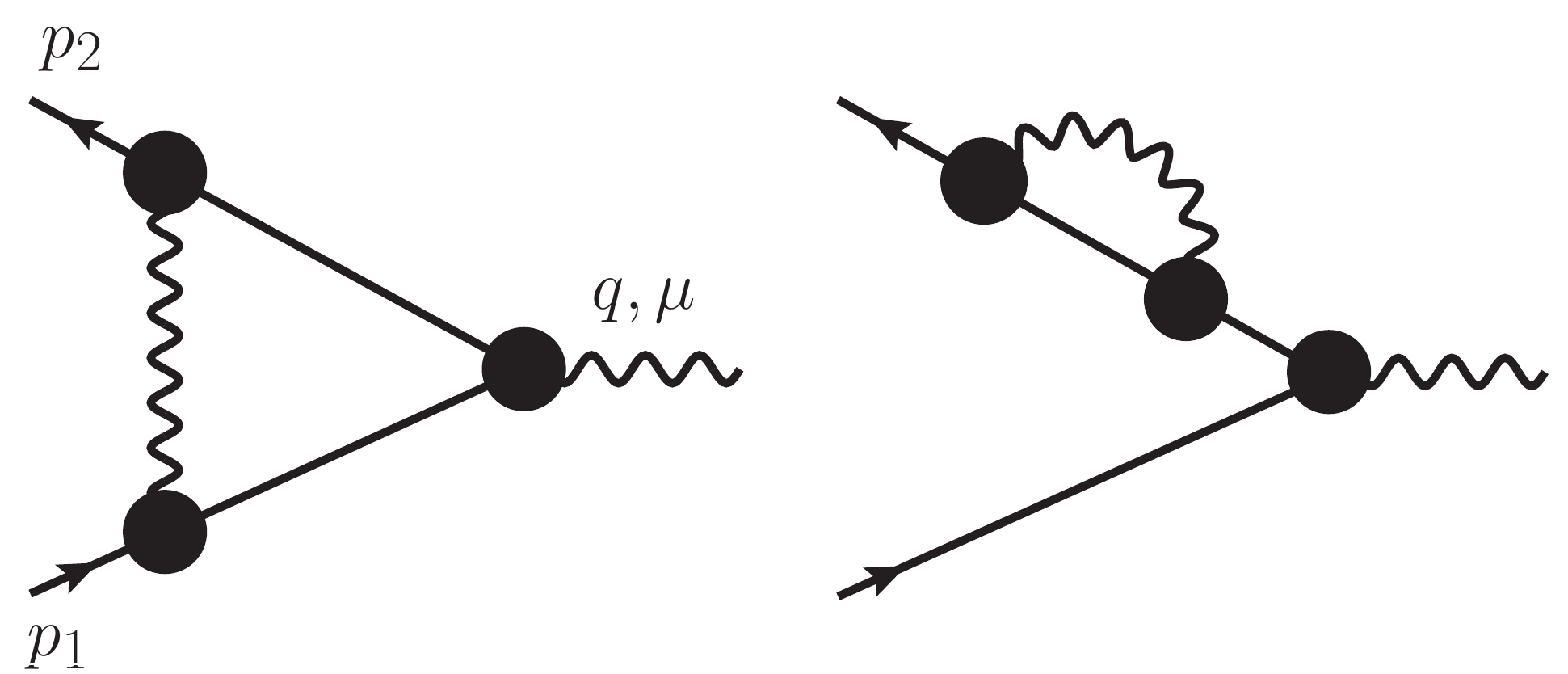}\caption{Photonic vertex and self-energy corrections with
  proton intermediate states.}
  \label{fig:pphloop}
\end{figure}

The contributions to $G_{E,M}^{\gamma{\rm-fin}}$ from the triangle diagram in Fig.~\ref{fig:pphloop} can be 
calculated numerically as double-integrals ${\alpha\over \pi}\int_0^\infty \!dx \int_{-1}^{+1} \!dz
\sqrt{1-z^2}\{[\dots]G_E(Q)+[\dots]G_M(Q)\}$ over cubic expressions in some given phenomenological form
factors $G_{E,M}$. The proton form factors  enter linearly in the external momentum transfer $Q$ and quadratically
in the loop-momentum $xM$ inside the square brackets. The
wave-function renormalization factor $Z_2$ from the self-energy diagram (see right part of
Fig.~\ref{fig:pphloop}) must also
be taken into account in order to ensure  that $G_E^\gamma(0)=0$. Using the representation of $Z_2$ in terms of
an integral over the proton form factors as written in Eq.~(5) of Ref.~\cite{Kaiser:2016tbf}, one
gets $Z_2= {\alpha\over \pi}\xi_{\rm ir}+Z_2^{\rm fin}$, where $Z_2^{\rm fin}$ depends weakly on the choice of
proton form factors $G_{E,M}$.

\begin{figure}[ht]
  \centering
  \includegraphics[width=0.4\textwidth]{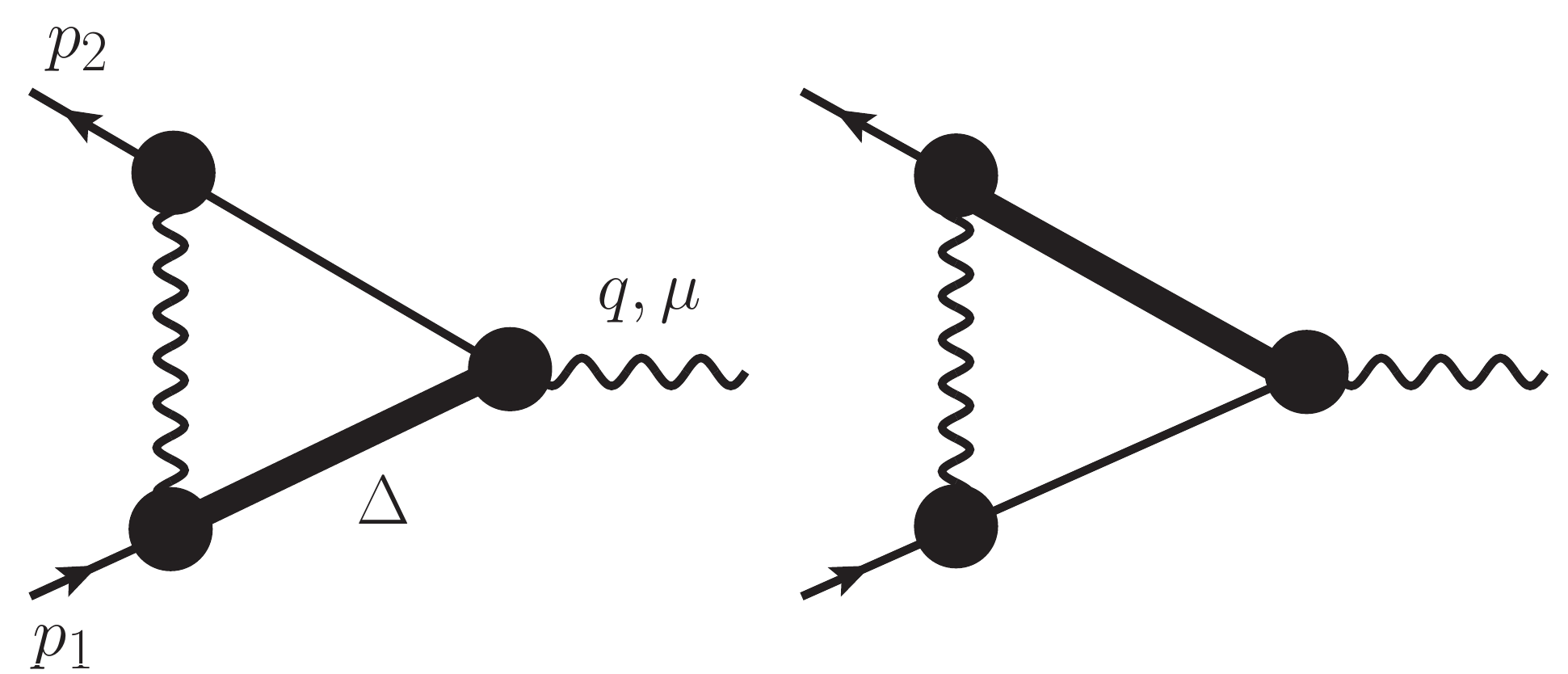}\hspace{1cm}\includegraphics[width=0.4\textwidth]{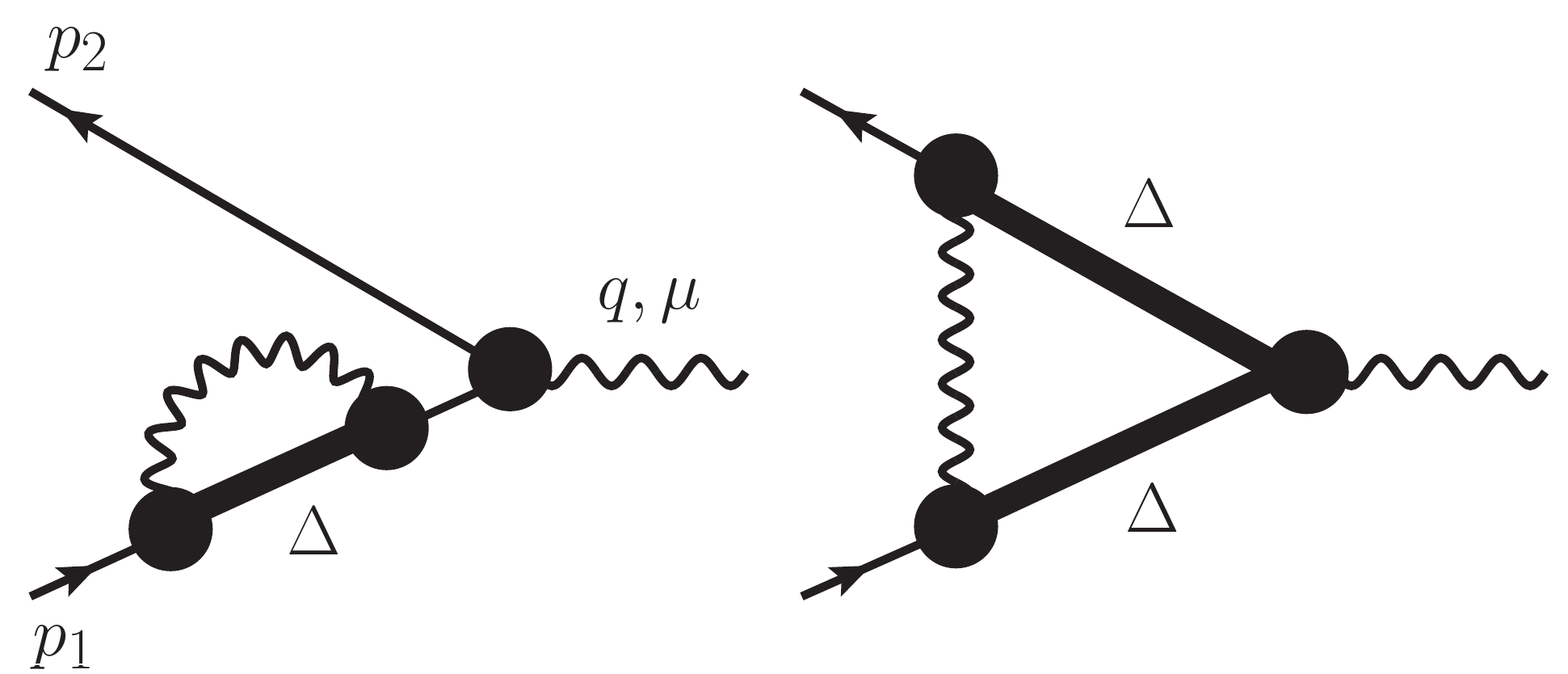}
  \caption{Photonic vertex and self-energy corrections with $\Delta^+(1232)$ intermediate states.}
   \label{fig:pphloopD}
\end{figure}

\begin{figure}[t]
\centering
\includegraphics[width=9.cm,clip]{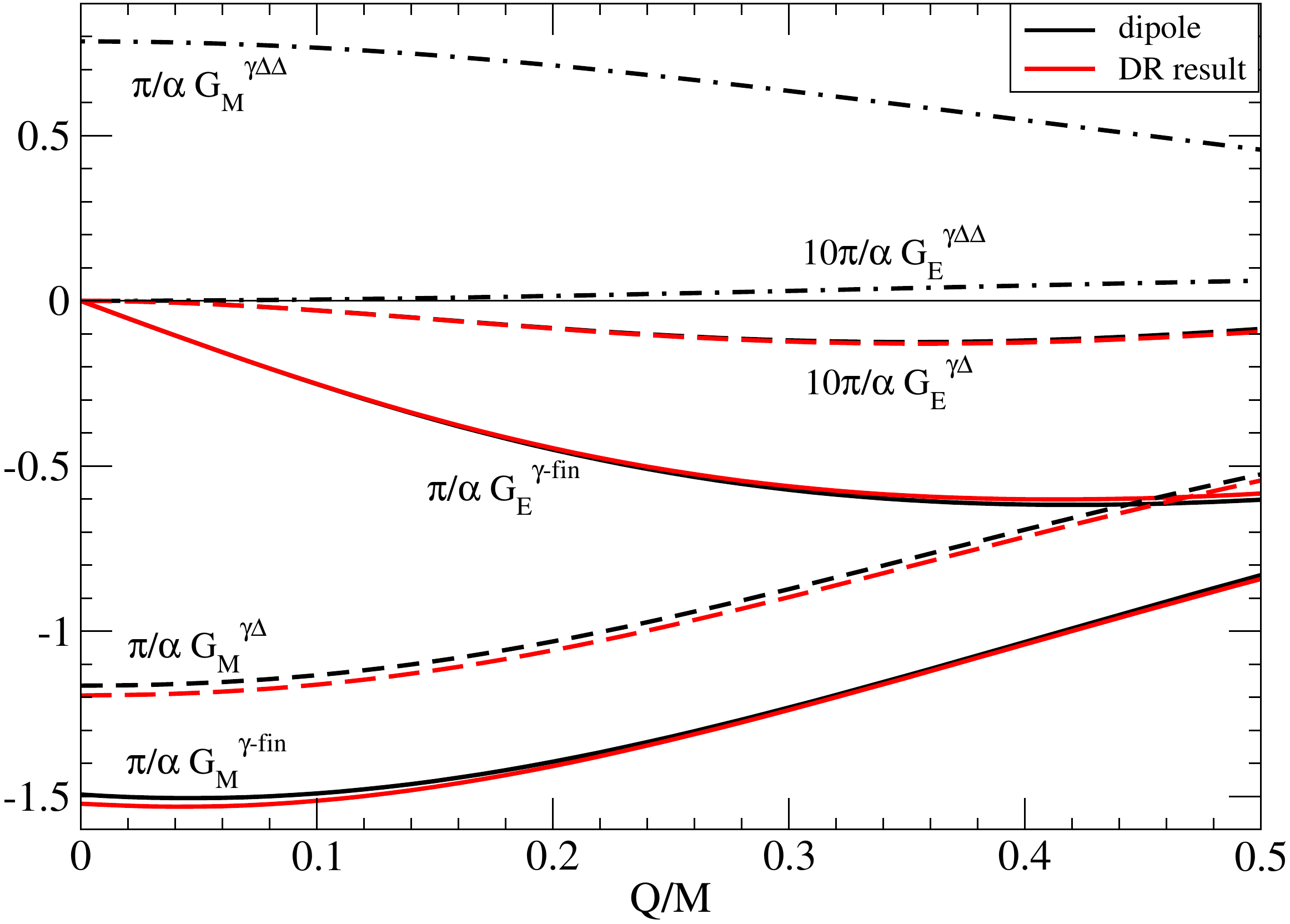}
\caption{Various contributions to the electric and magnetic photon-loop form factors of the proton. The
  black lines represent results obtained by using the dipole parametrization and the red lines show the
  calculations based on the form factors from~Ref.~\cite{Lin:2021xrc}.}
\label{fig:pphloops}
\end{figure}

In this work we model inelastic contributions to the photon-loop form factors of the proton by single and
double excitation of the $\Delta^+(1232)$ resonance, as shown in Fig.~\ref{fig:pphloopD}.  As we will see later, this is appropriate here. More sophisticated approaches could also be entertained, see e.g. the review
\cite{Arrington:2011dn}, but we refrain 
from doing  so here.  The Lorentz-covariant description of the spin-$3/2$
$\Delta$-isobar requires a Rarita-Schwinger spinor $\Psi_\alpha$. We use a minimal gauge-invariant form of
the $\Delta^+p\gamma$-vertex: 
\begin{equation}
\label{eq:DPg}
{ie\kappa^*\over\sqrt{6}M}\big(g^{\mu \alpha} \gamma\!\cdot\!q -\gamma^\mu q^\alpha\big)\gamma_5\,
G_\Delta(\sqrt{-q^2})\,,
\end{equation}
in terms of the transition magnetic moment $\kappa^*\simeq 5.0$ and a phenomenological transition form factor
\begin{equation}
\label{eq:Dff}
G_\Delta(Q) = \bigg(1+{Q^2 \over  \Lambda^2}\bigg)^{-2}\exp\bigg(-{Q^2\over 7\Lambda^2}\!\bigg)\,,
\end{equation}
with dipole mass $\Lambda = 843\,$MeV, as extracted from pion electroproduction in the $\Delta$-resonance
region~\cite{Albrecht:1971rv,Galster:1972rh,Burkert:1992yk}. Here, $q$ denotes the space-like four-momentum carried
by the virtual photon, so that $Q=\sqrt{-q^2}$. A commonly used form of the
Rarita-Schwinger propagator  (from index $\beta$ to  index $\alpha$) reads:
\begin{equation}
{i\over 3}\, {\gamma\!\cdot\!p +
M_\Delta \over M_\Delta^2-p^2}\bigg(3 g_{\alpha\beta} -\gamma_\alpha 
\gamma_\beta-{2p_\alpha p_\beta \over M_\Delta^2} +{p_\alpha 
  \gamma_\beta-\gamma_\alpha p_\beta\over M_\Delta} \bigg)\,,
\end{equation}
with $p$ the four-momentum of the propagating $\Delta$-isobar. The left two diagrams in Fig.~\ref{fig:pphloopD} provide equal contributions to $G_{E,M}^{\gamma\Delta}$, respectively,
and the condition $G_E^{\gamma\Delta}(0)=0$ follows immediately from the magnetic coupling vertex in Eq.~\eqref{eq:DPg}.
In order to evaluate the right diagram in Fig.~\ref{fig:pphloopD}, the $\Delta^+\Delta^+\gamma$-vertex is needed. It is naturally
obtained by gauging the kinetic term of the free Rarita-Schwinger Lagrangian \cite{Benmerrouche:1989uc} as:
\begin{equation}i e\big( -\gamma^\mu g_{\alpha\beta} + 
\gamma_{\alpha} g_\beta^\mu + \gamma_{\beta} g_\alpha^\mu - \gamma_\alpha
\gamma^\mu\gamma_\beta\big)\,
\end{equation}
with $\beta$ the incoming and $\alpha$ the outgoing (Rarita-Schwinger) index. Further, we  multiply this electric vertex
with a dipole form factor times a (squared) exponential function as in Eq.~\eqref{eq:Dff}. The
third diagram in Fig.~\ref{fig:pphloopD} contributes as $Z_2^{(\Delta)}G_{E,M}$, with the wavefunction renormalization
factor $Z_2^{(\Delta)}$ written in Eq.~(16) of Ref.~\cite{Kaiser:2016tbf}. It is worth to mention that only the
combination of both right diagrams ensures the condition $G_E^{\gamma\Delta\Delta}(0)=0$ for the photon-loop
induced electric form factor.  The various contributions to the photon-loop proton form factors
multiplied with $\pi/\alpha$ are shown
in Fig.~\ref{fig:pphloops}, for two choices of proton form factors, namely the well-known (simple) dipole form and
the most recent parametrization based on dispersion theory, that describes essentially all data in the time-like
and in the space-like regions~\cite{Lin:2021xrc}. We note that there are cancellations between the single- and
the double-$\Delta$ contributions to both form factors, and that only the $\Delta$ contribution to the magnetic 
form factor is sizeable (on the scale of $\alpha/\pi= 2.323 \!\cdot\! 10^{-3}$). 
Although the curve for $G_E^{\gamma-{\rm fin}}$ shown in Fig.~\ref{fig:pphloops} appears to drop
linearly for small $Q$, the underlying analytical expression is manifestly even under
$Q\to -Q$, and this is true for any form factor.

\section{Two-photon exchange box diagrams}
\label{sec:tpe}
We now consider the two-photon exchange correction incorporated in the $H_2$ term.
For a point-like proton with $G_E=G_M=1$ the ratio $H_2/H_0$ can be inferred from the calculation in
Sec.~3 of Ref.~\cite{Kaiser:2010zz} as:
\begin{equation}
  {H_2 \over H_0} = -{2t \over A'\otimes A'} {\rm Re}({\rm III}'\otimes A'+{\rm IV}'\otimes A') \,,
\end{equation}
where the change of sign is necessary, because there the case of equally
charged leptons of different masses has been considered. The pertinent one-photon exchange term is $A'\otimes A'=
2(s-1-r)^2+2st+t^2$ and ${\rm III}'\otimes A'$ is written in Eq.~(31) of ref.\cite{Kaiser:2010zz},
while ${\rm IV}'\otimes A'=-{\rm III}'\otimes A'|_{s\to u}$

\begin{figure}[t!]
\centering \includegraphics[width=9.cm,clip]{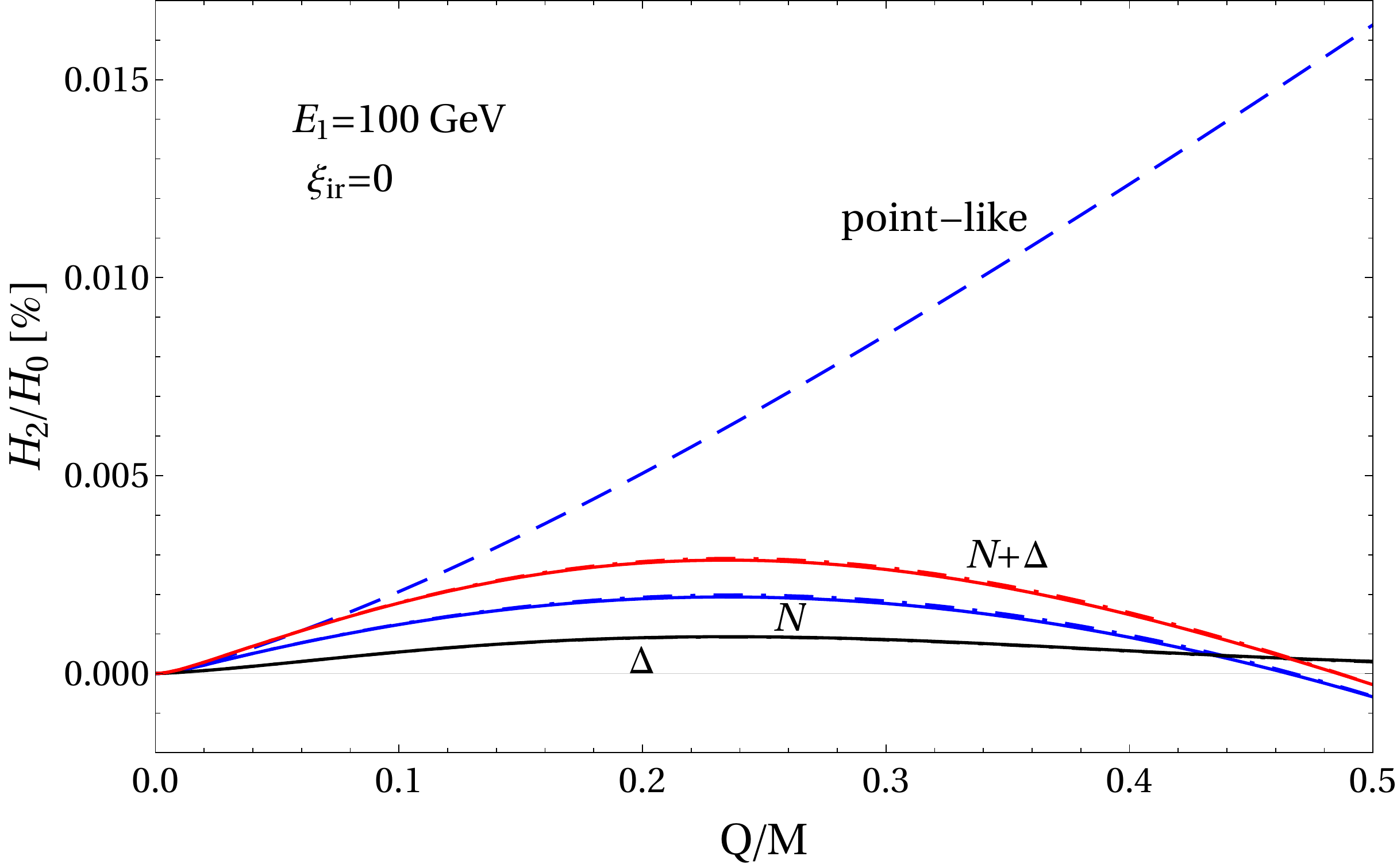}
\caption{The ratio $H_2/H_0$ for a point-like proton (dashed line) and for the physical proton
(lines labelled ``$N+\Delta$''). The elastic contribution and the inelastic contribution (modelled by single $\Delta$-isobar excitation) are shown separately  by the lines labelled ``$N$'' and ``$\Delta$''. Solid lines refer to the use of the form factors from Ref.~\cite{Lin:2021xrc} and dot-dashed lines are based on dipole form factors.}
\label{fig:tpe}
\end{figure}

For a structureless proton the ratio $H_2/H_0$ (setting $\xi_{\rm ir}=0$) at the AMBER kinematics
($E_1=100\,$GeV or $s\simeq 214$) is found to be rather small, starting from zero at $Q=0$ and increasing to
about $0.016\%$ at $Q=M/2=469\,$MeV, see Fig.~\ref{fig:tpe}. This small value results from a strong
cancellation between the planar and crossed $2\gamma$-exchange box graph, namely  $ 7.743\%-7.727\%$.
In the limit $Q\to 0$, where the cancelation becomes exact, the  contributions from the planar and crossed
box graph behave each as $\xi_{\rm ir}+\ln(Q/M)$ times an $(r,s)$-dependent factor of opposite sign.      
As also shown in  Fig.\,5, this cancellation is even more pronounced for the physical proton.
For the nucleon intermediate state, we use the Feynman graph formalism of Ref.~\cite{Lorenz:2014yda} and
the $\Delta$ intermediate state is evaluated using the dispersion relation approach of
Refs.~\cite{Blunden:2017nby,Tomalak:2014sva}.
The ratio $H_2/H_0$ does not exceed the value $0.003\%$ for the low momentum transfers considered here.

\section{Soft photon radiation}
\label{sec:softrad}

\begin{figure}[ht]\centering
\includegraphics[width=0.8\textwidth]{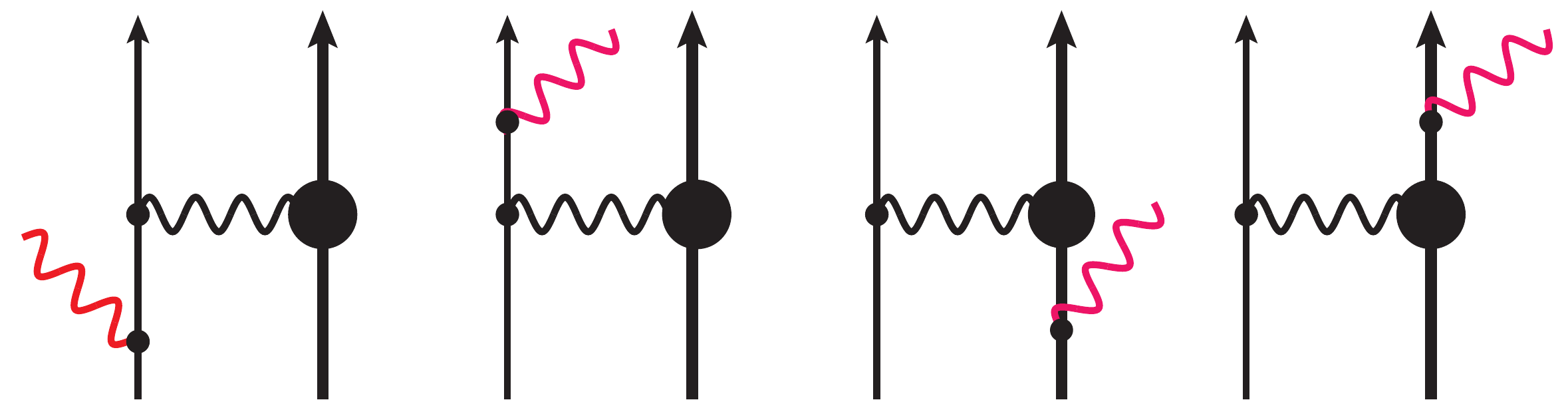}
\caption{Diagrams representing the four amplitudes for soft photon radiation. Thin and solid lines
  denote muons and protons, respectively, while the wiggly lines denote photons. The corresponding (point-like and structureful)
  vertices are depicted by small and large filled circles.}
\label{fig:soft}
\end{figure}

Without the inclusion of the soft photon radiation the treatment of radiative
corrections to $\mu^-p\to \mu^-p$ is incomplete or even meaningless.
When working at higher order in the electromagnetic coupling $e$, the muon and proton can radiate
a real photon with four-momentum $\ell$ and polarization vector $\epsilon$ in the initial or final state,
as shown in Fig.~\ref{fig:soft}. The corresponding soft amplitude reads:
\begin{equation}
e\bigg( {\epsilon\!\cdot \!k_1 \over \ell\!\cdot\! k_1}- {\epsilon\!\cdot\! k_2 \over \ell\!\cdot\! k_2}
-{\epsilon\!\cdot\! p_1 \over \ell\!\cdot \!p_1}+ {\epsilon\!\cdot\! p_2 \over \ell\!\cdot \!p_2}\bigg)\,,
\label{eq:softamp}
\end{equation} 
where the soft momentum $\ell$ is neglected in the numerator, and the accompanying $\mu^-p\to \mu^-p$
process is kept in the limit $\ell = 0$. Note that the signs in Eq.~\eqref{eq:softamp} indicate the charge
of the radiating particle and whether this emission happens in the initial or the final state. The soft
amplitude gets squared and summed over the two transversal polarizations by employing $\sum_{pol}
\epsilon^\mu \epsilon^\nu = - g^{\mu\nu}$, which leads to:
\begin{eqnarray}
&& 4\pi \alpha\bigg\{-{m_\mu^2 \over (\ell\!\cdot\! k_1)^2}-{m_\mu^2 \over (\ell\!\cdot\! k_2)^2}
-{M^2 \over (\ell\!\cdot\! p_1)^2} -{M^2 \over (\ell\!\cdot\!p_2)^2} \nonumber \\
&&  +{2k_1\!\cdot\!k_2 \over \ell\!\cdot\! k_1 \,\ell\!\cdot\! k_2}+{2p_1\!\cdot\!p_2 \over \ell\!\cdot\!p_1\, \ell\!\cdot\! p_2}+{2k_1\!\cdot\!p_1 \over \ell\!\cdot\! k_1\,
\ell\!\cdot\!p_1}+{2k_2\!\cdot\!p_2 \over \ell\!\cdot\! k_2 \,\ell\!\cdot\! p_2}-{2k_1\!\cdot\!p_2 \over \ell\!\cdot\! k_1 \,\ell\!\cdot\! p_2}-{2k_2\!\cdot\!p_1 \over
\ell\!\cdot\! k_2 \,\ell\!\cdot\! p_1}\bigg\}\,.
\label{eq:softampint}
\end{eqnarray}
In any experiment with finite energy resolution, the emission of additional soft photons with
$|\vec \ell\,|<\lambda$ is undetectable and thus gets subsumed in the cross section for the
elastic scattering process. The integrals of the ten terms in Eq.~\eqref{eq:softampint} over a (small) momentum
sphere $|\vec \ell\,|<\lambda$ are solved with the help of the following (infrared-regularized) master
integral: 
\begin{equation}
\int_0^\lambda \!\!d\ell {\ell^2 \over 2 \sqrt{m_\gamma^2+\ell^2}} \int_{-1}^{+1}\!\!dz {-1 \over \big(
E\sqrt{m_\gamma^2+\ell^2}- p \ell z\big)^2} = {1\over E^2-p^2}\bigg[ \ln {m_\gamma\over 2\lambda} +{E \over 2p}
\ln{E+p \over E-p}\bigg]\,.
\end{equation}
It applies directly to  the first four terms in Eq.~\eqref{eq:softampint}  with squares in the denominator,
whereas for the six terms with products in the denominator one makes use of the Feynman parametrization:
$(A B)^{-1} = \int_0^1\!dx [A x+B(1\!-\!x)]^{-2} $. Putting all pieces together, the correction factor
from soft photon radiation is given by the sum $\delta_{\rm soft}= \delta_{\rm soft}^{\rm (uni)}+ \delta_{\rm soft}^{\rm (cm)}$,
where the universal part reads:
\begin{eqnarray}
\delta_{\rm soft}^{\rm (uni)}&\!\!\!\!=\!\!\!\!& {4\alpha \over \pi} \Big( \ln{M\over 2 \lambda}-\xi_{\rm ir}\Big)
\bigg\{ 1+ {t-2r \over \sqrt{t^2-4r t}}\ln {\sqrt{4r-t}+\sqrt{-t} \over 2\sqrt{r}} +{t-2 \over \sqrt{t^2-4 t}}
\ln {\sqrt{4-t}+\sqrt{-t} \over 2}  \nonumber \\
&&  +{2(1+r-s)\over \sqrt{s-\rho_+} \sqrt{s-\rho_-}}\ln {  \sqrt{s-\rho_+}+ \sqrt{s-\rho_-}\over 2 r^{1/4}}
+{2(1+r-u)\over \sqrt{\rho_+-u} \sqrt{\rho_--u}}\ln {  \sqrt{\rho_+-u}+ \sqrt{\rho_--u}\over 2 r^{1/4}}\bigg\} ,
\nonumber \\
\label{eq:softuni}
\end{eqnarray}
with $\rho_\pm = 1+r \pm 2 \sqrt{r}$.  It cancels exactly 
the infrared divergences proportional to $\xi_{\rm ir}$ from the virtual photon-loops and the remainder depends
logarithmically on an infrared  cut-off $\lambda$ for undetected soft photon radiation. Note that the last
two terms cancel the infrared divergences from the two-photon exchange box diagrams. For
these the antisymmetry under $s \leftrightarrow u$ is manifest by setting $\sqrt{u-\rho_\pm }=i\sqrt{\rho_\pm-u}$,
$\sqrt{\rho_\pm-s }=i\sqrt{s-\rho_\pm}$ and taking eventually the real part of a logarithm.  In the limit
$r\to 0$ the sum of these two terms simplifies drastically to $\ln(1-u)-\ln(s-1)$. For $\mu^+ p\to \mu^+ p$
the last two terms in Eq.~(\ref{eq:softuni}) change sign, keeping the role of $s$ and $u$.

%At this point we have to note that the expression for the infrared-divergent term given in Eq.~(47) of
%Ref.~\cite{Tomalak:2018jak} is incorrect. The prefactor needs to be divided by $2$, and the term $(M+m)^2$ in
%the numerator and denominator of logarithms has to be replaced by $M^2+m^2$.      

The other part of $\delta_{\rm soft}$ is specific for assuming in the center-of-mass frame a small momentum
sphere $|\vec \ell\,|<\lambda$ for undetected soft bremsstrahlung: 
\begin{eqnarray}
\delta_{\rm soft}^{\rm (cm)}&\!\!\!\!=\!\!\!\!& {\alpha \over \pi}\Bigg\{{2\over \sqrt{P}} \bigg[(s-1+r) \ln{s-1+r 
+\sqrt{P}\over 2 \sqrt{s r}}+(s+1-r) \ln{s+1-r +\sqrt{P}\over 2 \sqrt{s }}\, \bigg]\nonumber \\
&& +\int_0^{1/2}\!\!\!dx\Bigg[{(t-2r)(s-1+r)\over [r-t x(1-x)]\sqrt{R_t}} \ln{s-1+r+\sqrt{R_t} \over s-1+r- \sqrt{R_t}}+ {(t-2)(s+1-r)\over [1-t  x(1-x)] \sqrt{R_t}} \ln{s+1-r+ \sqrt{R_t}
\over s+1-r-\sqrt{R_t}}  \,
\Bigg] \nonumber \\
&&+\int_0^1\!\!dx\Bigg[{(1+r-s)[s+(1-r)
(1-2x)]\over (1-2x)[s x(1-x)+(1-2x)(1-x-r x)] \sqrt{P}}
 \ln{s+(1-2x)(1-r+\sqrt{P})\over s+(1-2x)(1-r-\sqrt{P}) } \nonumber \\ && \qquad \qquad
+{(1+r-u)[s+(1-r)(1-2x)] \over  [1+(r-1)x-u x(1-x)] \sqrt{R_u}} 
\ln{s+(1-r)(1-2x)+\sqrt{R_u}\over s+ (1-r)(1-2x)-\sqrt{R_u}}\,\Bigg]\Bigg\}\,, 
\label{eq:softcm}
\end{eqnarray}
introducing the auxiliary polynomials $R_t=P+4s tx(1-x)$ and $R_u=P+4x(1-x)[s u-(1-r)^2]$,
with $P=s^2-2s(1+r)+(1-r)^2$. Although possible,
we refrain from solving the Feynman-parameter integrals in Eq.~(\ref{eq:softcm}) which lead to overly complicated
expressions involving squared logarithms and dilogarithms (see e.g. Eqs.(4.10)-(4.12)  in Ref.~\cite{Maximon:2000hm}).
For $\mu^+ p\to \mu^+ p$ the last term $\int_0^1\!dx\dots$ changes sign, keeping the role of $s$ and $u$.

\section{Results and discussion}
\label{sec:res}

\begin{figure}[t!]\centering
\includegraphics[width=0.45\textwidth]{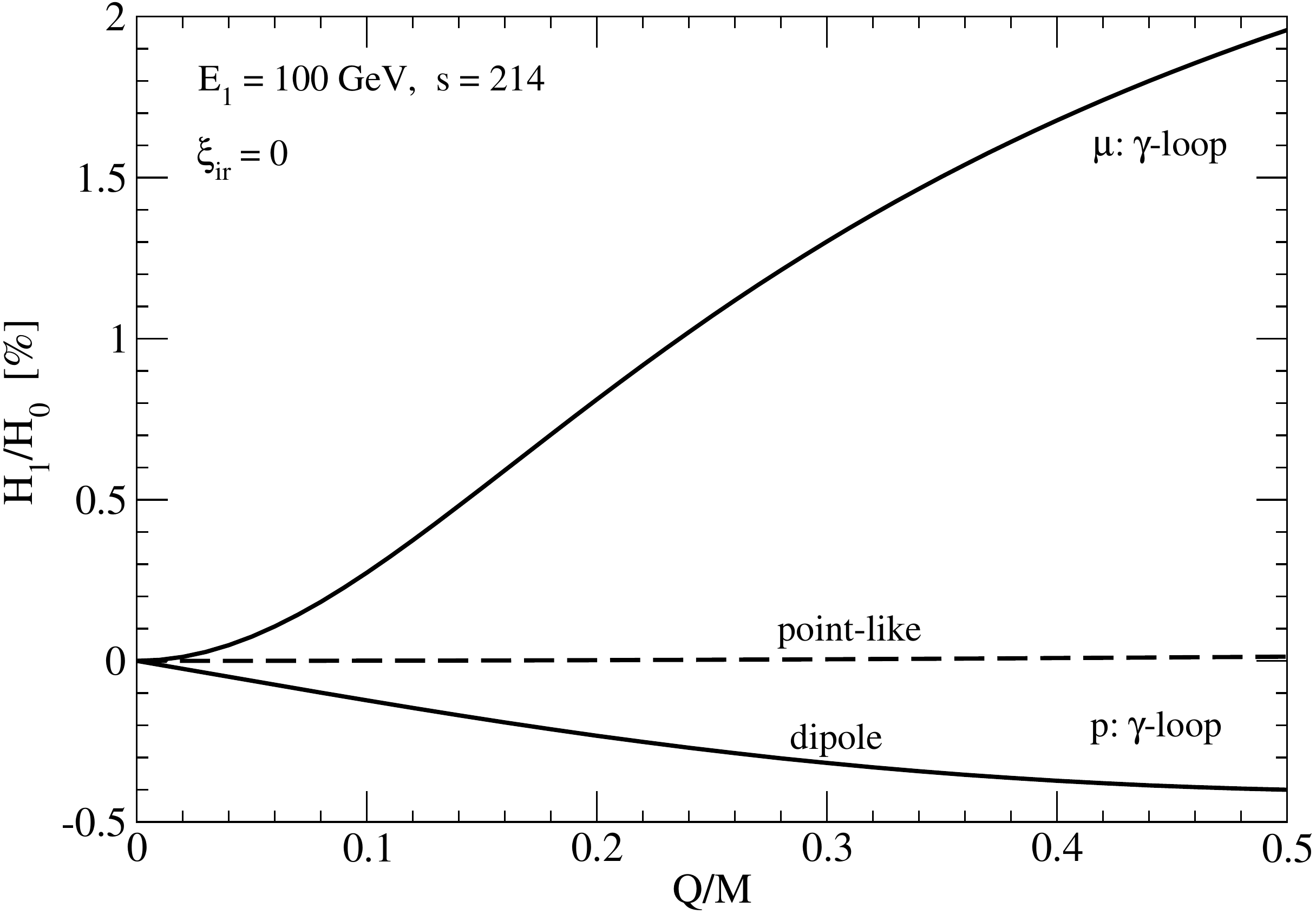}~~~
\includegraphics[width=0.45\textwidth]{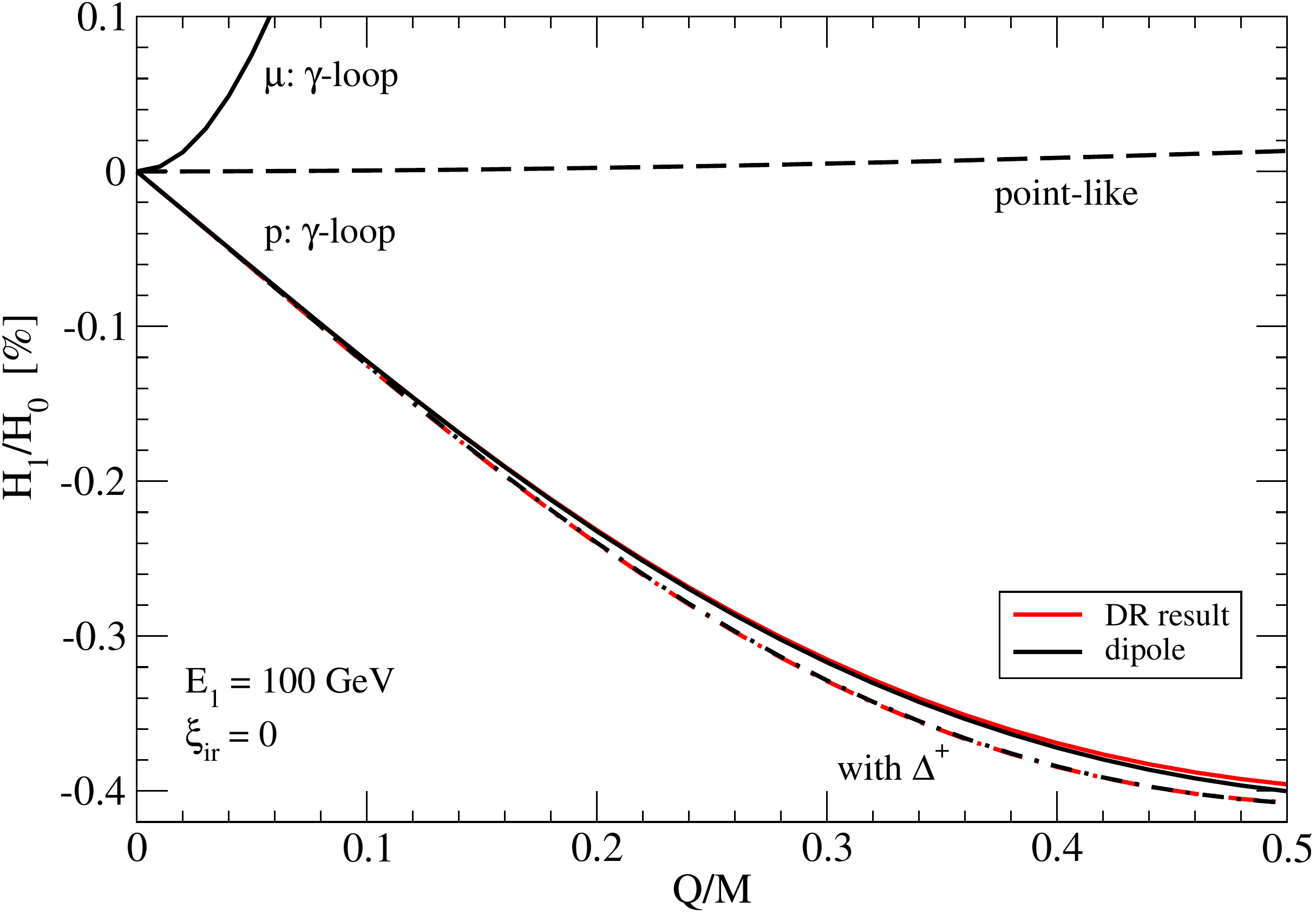}
\caption{Ratio $H_1/H_0$ due to electromagnetic vertex corrections separated into parts from the muon and the proton.
Left panel: Corrections from the photon-loop around the muon, the point-like proton and the composite proton using dipole form factors. Right panel: Zoom in on the proton contribution using dipole
form factors as well as the most recent form factors from dispersion theory~\cite{Lin:2021xrc}. The solid
and dot-dashed lines refer to the elastic and elastic plus inelastic contribution.}
\label{fig:H1overH0}
\end{figure}

We can now put the pieces together and discuss the radiative corrections to muon-proton scattering. First, it is interesting to see how much the radiative corrections to $\mu^-p\to \mu^-p$ scattering
(at a beam energy of $E_1=100\,$GeV) are affected by the underlying proton structure. For that purpose we
make comparisons with a structureless proton, corresponding to $G_{E,M}=1$. Returning to Eq.~\eqref{eq:dxs},
one recognizes that the radiative corrections (i.e. changes of cross section ratios) from vacuum polarization and soft photon bremstrahlung are
the same for a structureful and structureless proton. Concerning the ratio $H_1/H_0$, one finds that this feature still holds for the muonic part written in the first line in  Eq.~\eqref{eq:H1}. To high accuracy
this ratio is equal to $2F_1^\gamma$, because the other muon form factor $F_2^\gamma$ is suppressed by a factor
of the  muon mass squared, and it enters with a further suppression factor $t$. Interestingly, the situation is
different for the vertex corrections at the proton, described by $G_{E,M}^\gamma$ in the second line of
Eq.~\eqref{eq:H1}. As shown in the left panel of Fig.~\ref{fig:H1overH0},
for a structureless proton\footnote{For a point-like proton one has $G_E^\gamma = F_1^\gamma+t F_2^\gamma/4$
  and $G_M^\gamma = F_1^\gamma+F_2^\gamma$ with $F_{1,2}^\gamma$ as in Eqs.~(\ref{eq:F1},\ref{eq:F2}) setting $r=1$.}
and  one finds that this part of the ratio $H_1/H_0$ grows from zero to
$1.33\cdot 10^{-4}$ in the momentum transfer region $0<Q<M/2$. If these vertex corrections, specified by
$G_{E,M}^{\gamma-{\rm fin}}$,  are evaluated with phenomenological form factors, the corresponding
radiative correction $H_1/H_0$ is negative and reaches more significant values of about $-0.4\%$. Further effects from
inelastic contributions (modelled here by $\Delta^+(1232)$-resonance excitations) turn out to be of
magnitude $10^{-4}$ and are thus not relevant in view of the experimental accuracy, see the right
panel of  Fig.~\ref{fig:H1overH0}.

\begin{figure}[t!]\centering
\includegraphics[width=0.8\textwidth]{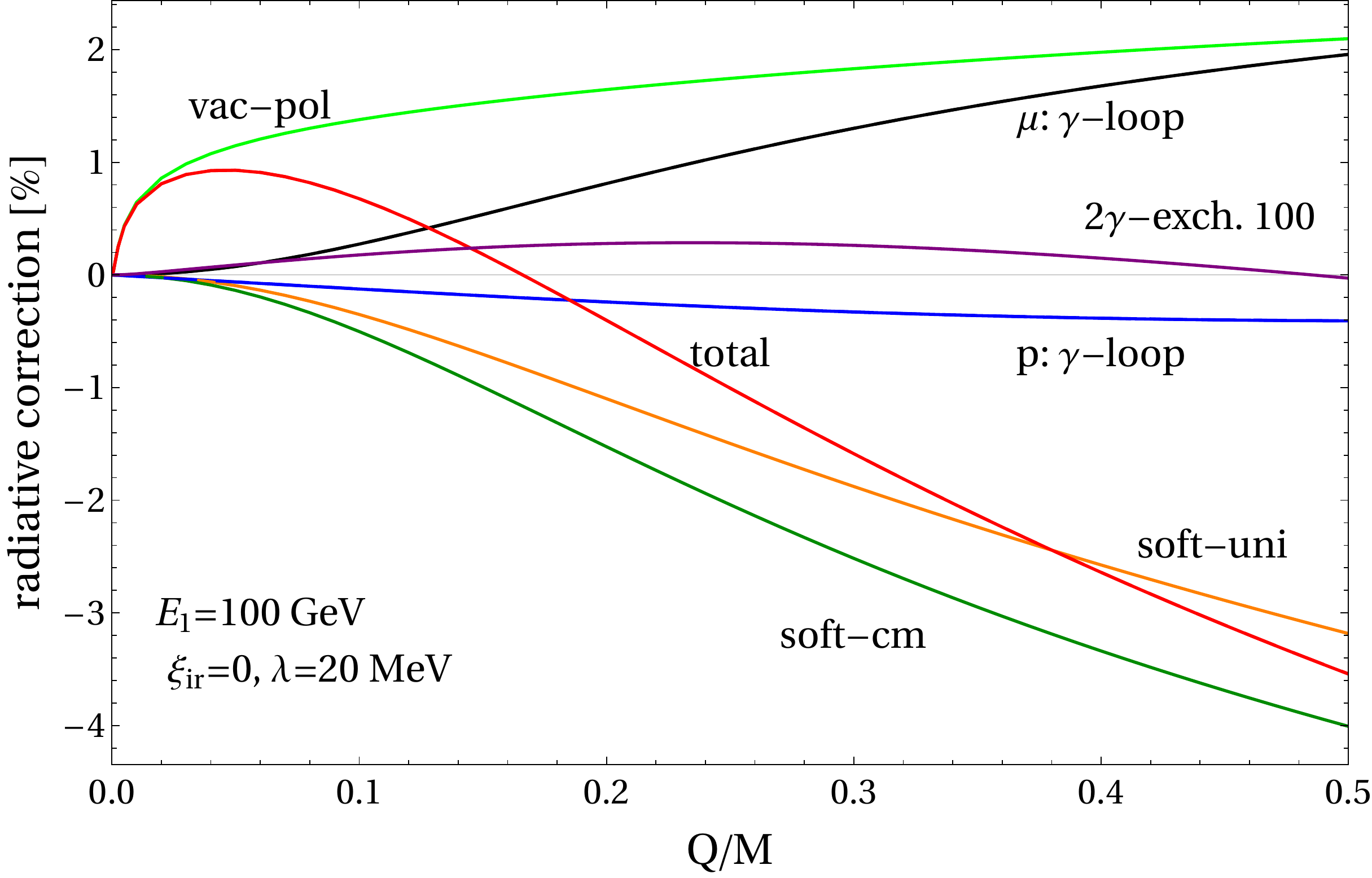}
\caption{Radiative corrections for the AMBER kinematics with $E_1=100\,$GeV. The individual
  radiative corrections from vacuum polarization, the virtual photon-loops and soft bremsstrahlung are
  shown together with their sum, see the solid line labelled ``total''.}
\label{fig:radcorr100}
\end{figure}

In Fig.~\ref{fig:radcorr100} we show the radiative corrections from all discussed sources (of order $\alpha/\pi$)
for the planned AMBER experiment with a muon beam energy of $E_1 =100\,$GeV
and an assumed infrared cutoff of $\lambda =20\,$MeV, corresponding to an energy resolution that limits the detection of photons  with an energy below $20\,$MeV. We set $\xi_{\rm ir} =0$ in order to have all individual contributions independent of the regulator mass $m_\gamma$. At small momentum transfers $Q/M \lesssim 0.06$, vacuum polarization is the most dominant effect, because it is driven by the electron mass scale.
After that, the soft-photon radiation takes over, with a sizeable contribution (of $2\%$) from the photon-loop form factor $2F_1^{\gamma-{\rm fin}}$, involving the muon mass scale, at the upper end of the
momentum transfers considered here. The negative photon-loop form factor contribution from the proton
stays below $0.4\%$ in magnitude, and the two-photon exchange correction of maximal size $0.3\cdot 10^{-4}$ can essentially be neglected.

\begin{figure}[t!]\centering
\includegraphics[width=0.8\textwidth]{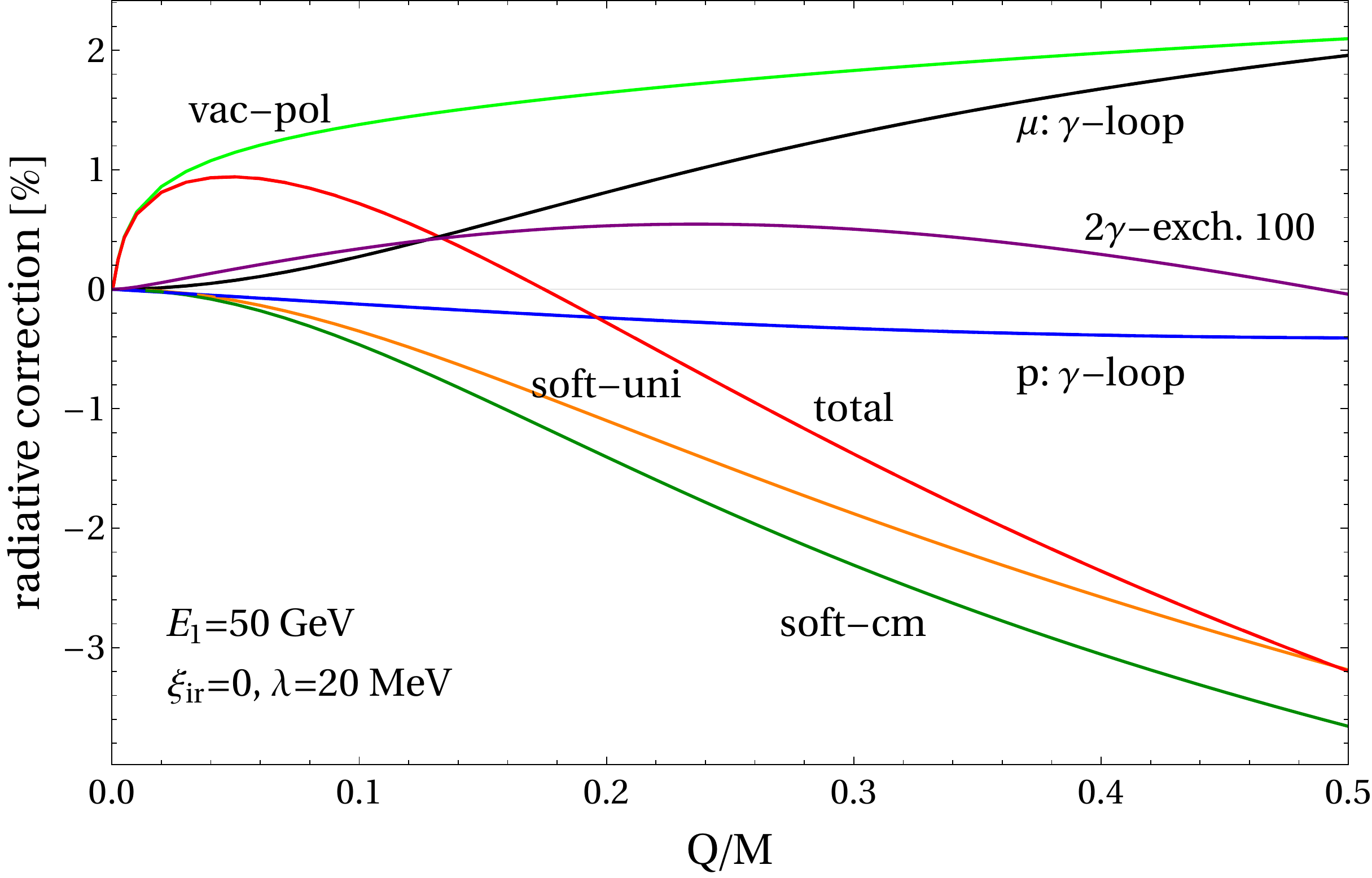}
\caption{Radiative corrections for the AMBER kinematics with $E_1=50\,$GeV. Same notation as
  in Fig.~\ref{fig:radcorr100}.}
\label{fig:radcorr50}
\end{figure}

As a variation  we show in Fig.~\ref{fig:radcorr50} the radiative corrections for a smaller muon beam energy of $E_1 =50\,$GeV (as it is also planned by AMBER) and the same infrared cutoff of $\lambda =20\,$MeV. In comparison to the  case with $E_1 =100\,$GeV, one observes an increase of the two-photon exchange correction and some slight changes in the soft photon components, but the overall pattern is nearly the same. 

\section{Summary and outlook}
\label{sec:summ}

In this  work we have systematically calculated the radiative corrections of
order $\alpha/\pi $ to elastic muon-proton scattering. These corrections consist of vacuum polarization,
soft photon radiation, photon-loop form factors of the muon and of the proton, and two-photon exchange corrections.
The first three components turn out to be universal in the sense that the structure of the proton
(as encoded in the electric and magnetic form factors $G_{E,M}$)  drops out in the respective ratios to the
Born cross section. The photon-loop induced vertex corrections at the proton give rise to additional form
factors $G_{E,M}^\gamma\sim \alpha/\pi$, whose infrared finite pieces can be calculated with sufficient accuracy.
The elastic contribution to  $G_{E,M}^\gamma$ from the proton intermediate state is almost independent of the
input form factors into the pertinent triangle diagram, while inelastic contributions (modelled here by
excitation of the low-lying $\Delta(1232)$-resonance) play numerically no role for the relevant ratio $H_1/H_0$.
The same is even more true for the two-photon exchange, whose relative effect measured by the ratio $H_2/H_0$
stays well below $10^{-4}$ in the small momentum transfer region $Q<400\,$MeV. Therefore, the aspects
of proton structure that enter the virtual radiative corrections do not limit the precision of extracting the
form factors $G_{E,M}$, and finally the proton radius  $r_p$, accurately from elastic muon-proton scattering   

With the complete radiative corrections of the size of a few percent and aiming for permille accuracy,
a prominent role is played by the soft photon radiation. The treatment of undetected soft
photon bremsstahlung in this work in terms of a small momentum sphere $|\vec \ell\,|<\lambda$ in the
center-of-mass frame corresponds to an idealized experiment.  In a real experiment this
region in  phase space has a more complicated structure with smooth edges due to varying detector acceptances
and other effects. By computing the fivefold differential cross section
$d^5\sigma/(d\Omega_\mu d\Omega_\gamma d\omega_\gamma)$ for the process $ \mu^\mp  p \to \mu^\mp p \gamma$ at tree-level,
and integrating it over the experimentally ``blind regions'' (of course, now with exclusion of the small
momentum sphere $|\vec \ell\,|<\lambda$) the treatment of undetected  soft and hard photon bremsstahlung
can be tailored to the specific experimental conditions. At the same time this
fivefold differential cross section can be used for experimental verification of its spectral and
angular distributions in regions where additional photons are
detectable. Work along these lines in collaboration with members of the AMBER collaboration is in progress.

\section*{Acknowledgements}

We gratefully acknowledge funding by  the Deutsche Forschungsgemeinschaft
(DFG, German Research Foundation) and the NSFC through the funds provided  to  the  Sino-German
Collaborative  Research  Center  TRR110  ``Symmetries  and  the  Emergence  of  Structure in  QCD''
(DFG  Project  ID 196253076  -  TRR  110,  NSFC Grant  No.  12070131001),
the Chinese Academy of Sciences (CAS) President's International Fellowship Initiative (PIFI)
(Grant No. 2018DM0034), Volkswagen Stiftung  (Grant  No.  93562),
the European Research Council (ERC) under the European Union's Horizon 2020 research and
innovation programme (grant agreement No. 101018170).
Further support by the DFG (Project ID 491111487) is  acknowledged.

%\newpage

\end{document}